# Single-shot multi-level all-optical magnetization switching mediated by spin-polarized hot electron transport

S. Iihama[1,2], Y. Xu[1], M. Deb[1], G. Malinowski[1], M. Hehn[1], J. Gorchon[1], E. E. Fullerton[1,3], and S. Mangin[1*]

* stephane.mangin@univ-lorraine.fr

[1] *Institut Jean Lamour, UMR CNRS 7198, Universite de Lorraine, Nancy, France*
[2] *National Institute of Advanced Industrial Science and Technology (AIST), Tsukuba, Japan*
[3] *Center for Memory and Recording Research, University of California San Diego USA*

**All-optical ultrafast magnetization switching in magnetic material thin film without the assistance of an applied external magnetic field is being explored for future ultrafast and energy-efficient magnetic storage and memories. It has been shown that femto-second light pulses induce magnetization reversal in a large variety of magnetic materials. However, so far, only GdFeCo-based ferrimagnetic thin films exhibit magnetization switching via a *single* optical pulse. Here we demonstrate the single-pulse switching of Co/Pt multilayers within a magnetic spin-valve structure ([Co/Pt] / Cu / GdFeCo) and further show that the four possible magnetic configurations of the spin valve can be accessed using a sequence of *single* femto-second light pulses. Our experimental study reveals that the magnetization final state of the ferromagnetic [Co/Pt] layer is determined by spin-polarized hot electrons generated by the light pulse interactions with the GdFeCo layer. This work provides a new approach to deterministically switch ferromagnetic layers and a pathway to engineering materials for opto-magnetic multi-bit recording.**

The possibilities of deterministically manipulating magnetization solely with ultra-short light pulses have attracted a growing attention over the past ten years leading to numerous ultra-fast and low-energy data storage concepts [1-9]. In 2007, all-optical switching (AOS) of magnetization using femtosecond (fs) laser pulses in GdFeCo ferrimagnet was first discovered [1]. Later on, it was shown that the magnetization in GdFeCo can be switched by a *single* fs-laser pulse independently of the light helicity [2,3] referred to as all-optical helicity-independent switching (AO-HIS) and has only been observed for GdFeCo-based materials. The AO-HIS has been described by a thermal-driven switching mechanism attributed to the transient ferromagnetic like states and the transfer of angular momentum between Gd sub-lattice and FeCo sub-lattice [2,3,17-20]. Very recently, this type of switching has not only been observed in the case of light pulses but also for electron pulses [7-9].

In contrast to AO-HIS, in the case of all-optical helicity-dependent switching (AO-HDS), the final state of magnetization is determined by the circular polarization of the light. AO-HDS has been observed for a large variety of magnetic material such as ferrimagnetic alloy, ferrimagnetic multilayer, ferromagnet thin films and granular recording media [4,5,10-15]. However, so far multiple pulses are necessary to fully deterministically switch the magnetization for AO-HDS [10, 16]. The use of single-pulse switching would be interesting because it is ultra-fast and energy-efficient, however restriction to Gd-based materials limits potential spintronic devices application. Furthermore, in order to move toward ultrafast-spintronic applications, one needs to study and understand the fundamental mechanism not only for single layers as it has been done in most study so far but also in more complex structures like spin valve structures, a key building block of modern spintronics. Selective magnetization switching in spin-valve structures or more complex heterostructures will enable multi-level magnetic storage and memories [21-23].

Here we demonstrate that the four possible magnetic configurations of a magnetic spin-valve structure ([Co/Pt] / Cu / GdFeCo) shown schematically in Fig. 1, where both layers are magnetically decoupled, can be accessed using a sequence of *single* femto-second light pulses. We show that a single laser pulse is able to switch the magnetization of either the GdFeCo layer alone or the magnetizations of both GdFeCo and [Co/Pt] layers, depending on the optical pulse intensity. We attribute this magnetic configuration control of the multilayer to, in part, a result of the ultrafast magnetization dynamics in spin-valve structure as well as ultrafast non-local transfer of angular momentum between layers [24-29]. Indeed, ultrafast quenching of magnetization in ferromagnetic or ferrimagnetic layers creates spin-polarized hot electrons that propagate in the metallic spacer layer and transfer the angular momentum to the other magnetic layer. We believe the switching of the [Co/Pt] layer results from a combination of optical excitation and the transfer of spin-polarized hot electron currents generated via the demagnetization of the GdFeCo layers.

*All Optical Multibit recording in GdFeCo / Cu / [Co/Pt] spin-valve structure*

Schematic illustration of the Ta(5nm)/Gd$_{23.3}$(FeCo)$_{76.7}$(5nm)/Cu(9.3nm)/[Co(0.6nm)/Pt(1.0nm)]$_4$/Ta(5nm)/Glass substrate spin valve structure namely GdFeCo/Cu /[Co/Pt] is shown in Fig. 1a. The GdFeCo and [Co/Pt] layers exhibit perpendicular magnetic anisotropy (PMA) and are magnetically separated by a 9.3-nm continuous Cu layer. The ferrimagnetic GdFeCo layer is FeCo rich at room temperature for this composition as the net magnetization of the alloy is parallel to the magnetization of FeCo sublattice and antiparallel to the magnetization of the Gd sublattice. Using magnetic fields, four remanent magnetic configurations can be reached: Two configurations where the magnetization of the two layers are parallel P+ and P- when the two magnetizations are along the positive and negative field direction, respectively, and two configurations where the two magnetizations are antiparallel AP+ and AP- when the magnetization of the [Co/Pt] is along the positive and negative field direction, respectively. Figure 1b shows the magneto-optic Kerr effect (MOKE) signal ($\theta_K$) as a function of the applied magnetic field (*H*) applied perpendicular to film plane. The reversal that appears at low magnetic field is attributed to the GdFeCo reversal. This could be confirmed by the magnetization curves (See supplementary information, sec A). Minor loops in blue and red in Figure 1.b show that the two layers are magnetically decoupled (i.e. the minor loop is centered about zero applied field).

In Fig. 2 we show how the four remanent magnetic configurations (P+, AP+, AP- and P-) can be accessed using *single* 35-fs light pulses under zero applied field. Starting with a saturated sample in the P- state (left most configuration in Fig. 2) one can switch both magnetic layers to the corresponding P+ state by sending a relatively intense pulse of 0.5 µJ where a region of roughly 30 µm in diameter is reversed. Then exposing the sample to a moderate light pulse of 0.2 µJ only triggers the reversal of the GdFeCo magnetization putting the sample into the AP+ state. A second 0.2-µJ pulse again reverses the GdFeCo layer returning the system to the P+ state. Applying a 0.5-µJ then reverses both magnetic layers back to the original P- state. Finally, a 0.2-µJ pulse again reverses the GdFeCo layer yielding the AP- state. Somewhat surprisingly if the AP+ or AP- state is excited by a 0.5-µJ pulse only the GdFeCo layer is reversed and the [Co/Pt] layer remains unperturbed. To summarize, 0.5-µJ pulses switch both layers from P+ to P- and back again while 0.2 µJ pulses are sufficient to only switch the GdFeCo layer leaving the [Co/Pt] layer unperturbed. The P+ to P- and P-to-P+ transitions clearly shows that the [Co/Pt] layer can be switched by a single 35 fs pulse.

***Energy and radius dependence of the single pulse on the magnetization switching***

In order to get insights on the energy and configurational changes of the switching of the [Co/Pt]/Cu/GdFeCo structures, we have studied the influence of the pump energy within the beam spot. Indeed, since the beam intensity follows a Gaussian profile, we can explore, with a single pulse, a range of energies. In Fig. 3a starting from an initial state in the P+ configuration, 35-fs single pulses of a given pump energy irradiated the sample consecutively with a typical time between two pulses of 2 seconds. The same experiment has been reproduced for 8 different pump energies ranging from 0.08 μJ to 0.36 μJ (three energies are shown in Fig. 3a). The spatial distribution of the magnetic configuration has been measured after each pulse. From Fig. 3a we can define up to four threshold radii ($r_i$) measured from the center of the optical excitation region whose values depend on the pump energy and allow us to distinguish five different responses to the initial magnetic configuration and light interaction.

The radius $r_1$ corresponds to the threshold for the onset of AO-HIS of the GdFeCo layer while leaving the [Co/Pt] layer unperturbed. Thus for $r>r_1$ the magnetization is unperturbed by the light while for $r<r_1$ the GdFeCo layer reverses its magnetization with each pulse from P+ to AP+ or AP+ to P+. This can be seen clearly for 0.08 μJ pulses in Fig. 3a where only within the radius $r=r_1$ there is a switching of the GdFeCo.

For increasing pulse energy of 0.2 μJ pulses (Fig. 3a) two new radii appear. With each pulse the outer ring corresponds to $r_1$ where only the GdFeCo reverses. Near the center of the beam there is a region for $r<r_3$ where both the GdFeCo and [Co/Pt] layers reverse from P+ to P- and then back with each pulse. There is a second radius, **r2**, where starting from P+ the GdFeCo magnetization is switched between $r_1>r>r_3$ to the AP+ configuration after the first pulse. Then for subsequent pulses there is a narrow region **r₂>r>r₃** where the light has no effect with further pulses. For yet higher pulse energy such as 0.32 µJ as shown in Fig. 3 where for r<**r₄** the energy density is sufficiently high that both layers demagnetize forming random domain states.

From images like those shown in Fig. 3a and corresponding line scans that quantify the magnetic contrast both the magnetic configurations and corresponding threshold radii can be determined and are shown in Fig. 3b for various pulse energies. The solid lines in Fig. 3d are fits assuming a Gaussian spatial distribution of the pump energy, *i.e.*, $F_\text{p}^i = 2E_\text{p}/(\pi w_0^2) \cdot \exp(-2r_i^2/w_0^2)$, where $F_\text{p}^i$, $r_i$ and $2w_0$ are threshold fluence, threshold radius, and $1/e^2$ spot size, respectively. From the fitting of our experimental data (from Fig. 3b), we are able to extract four (incident) threshold fluences, $F_\text{p}^1$= 3.1 mJ/cm², $F_\text{p}^2$ = 8.5 mJ/cm² $F_\text{p}^3$ = 11.8mJ/cm² and $F_\text{p}^4$ = 28.4 mJ/cm² that correspond to radii **r₁** through **r₄**. We note that the range of fluences for which single shot switching is observed is the highest ever reported. In our films switching fluences cover a whole decade (3.1 – 28.4 mJ/cm2), whereas in GdFeCo single layers doubling the AOS threshold fluence would lead to multidomain creation [30, 31]. We believe this behavior might be related to the higher Curie temperature of the [Co/Pt] layer and the dynamic coupling between the layers.

***Discussion of the mechanism of single shot switching of different regions***

For low fluences (**r₁>r>r₂**) we observe that only the GdFeCo switches with each pulse as has been observed as in single GdFeCo films. At these fluences the [Co/Pt] layer is not sufficiently excited to be perturbed. This behavior is expected from the extensive literature of AO-HIS on GdFeCo films [32]. The narrow region $r_2>r>r_3$ where once the sample is in the antiparallel configuration no changes occur after next pulse indicates some subtle interplay between layers that prefers the antiparallel configuration. However, we do not have a detailed explanation of this interaction at present.

We will focus much of the remainder of the discussion on the region $r_3>r>r_4$ where we observe the simultaneous reversal of the GdFeCo and [Co/Pt] layers by single pulses. We believe that in this region, GdFeCo is switching by itself via the AO-HIS but in addition there is single shot deterministic reversal of the ferromagnetic layer through a combination of optical excitation and a dynamic coupling mechanism between the layers. We will discuss possible coupling mechanisms below and discuss additional experiments to test the validity of these mechanisms.

The first possible explanation is the presence of a static exchange coupling, which has been reported to be at the origin of the single shot AOS of a [Co/Pt] bilayer coupled to a GdFeCo layer [34]. In this study the final state of the [Co/Pt] layer is determined by the sign of the interlayer coupling. For our samples, the minor loops of the GdFeCo magnetization reversal show no measurable exchange coupling between the two magnetic layers (Fig. 1). The fact that we can independently switch the GdFeCo layer and the energy required for AO-HIS of the GdFeCo layer from P+ to AP+ and from AP+ to P+ are identical, also suggest there is no exchange coupling via the Cu interlayer present in our sample. Moreover we observed this single-shot AO-HIS of the GdFeCo and [Co/Pt] layers for up to a 30 nm-thick Cu spacer layer (See supplementary information, sec C) where no exchange coupling is expected.

A second possible explanation could be the presence of some dipolar coupling. While there are no dipolar fields arising from an ideal uniformly magnetized film, there are dipolar fields generated from domain states and inhomogeneous magnetization. For circular domains such are shown in Fig. 2 the dipolar fields are strongest at the boundary and may be responsible for the narrow region $r_2>r>r_3$. To explore the role of dipolar coupling we studied samples with a different GdFeCo alloy concentration such that the net magnetization of the ferrimagnetic GdFeCo would be in the direction of the transition-metal sublattice (*i.e.* "FeCo-rich") or the rare-earth sublattice (*i.e.* "Gd-rich"). Figure 4 shows AOS in spin-valve structure of the FeCo-rich sample (Fig. 4a) and the Gd-rich (Fig. 4b) GdFeCo. Since MOKE measurements are mostly sensitive to FeCo sublattice moment (as opposed to the total moment), magnetic contrast values in antiparallel configuration of the magnetization are higher than that in parallel configuration in Gd-rich GdFeCo as shown in Fig. 4b. Single-shot AO-HIS of the GdFeCo and [Co/Pt] layers is observed for both GdFeCo concentrations. However the final magnetic state reached by the GdFeCo and [Co/Pt] layers corresponds to a parallel alignment of the FeCo sublattice and the Co/Pt magnetizations independently of the GdFeCo concentration and net magnetization as shown schematically in Figs. 4c and d. Thus dipolar interactions (at room temperature) are unlikely to explain the final state of the magnetization of the GdFeCo and [Co/Pt] layers.

Having ruled out indirect and dipolar coupling we believe the final state of the ferromagnetic [Co/Pt] layer is determined by spin-polarized hot electrons generated by the light pulse interactions with the GdFeCo layer in addition to local heating of the [Co/Pt] layer. To explain the single pulse AO-HIS of the [Co/Pt] layer, we assumed that the ultrafast laser heating after interacting with the GdFeCo layer is generating a hot electron spin polarized current (i.e. superdiffusive spin currents) parallel to Gd moment which will ultimately transfer its angular momentum to the [Co/Pt] layer and in combination with optical excitation determines the final state of magnetization (schematically shown in Fig. 4c and 4d.). Recent experiments on spin currents generated by the interaction between a light pulse and a GdFeCo layer [29] observed, as a function of time after optical excitation, first positive and then negative spin generations in the conduction band. The authors attribute the positive part of the spin generation to demagnetization of the transition-metal sublattice, and estimate the majority of the negative spin is coming from slower demagnetization of the Gd sublattice. They also suggest a potential contribution to the negative spin current to the spin-dependent Seebeck effect. Both contributions result in the longer time spin currents being parallel to the Gd sublattice, as shown in Figs. 4c and 4d, that will be transferred and absorbed by the [Co/Pt] layer. If the Co/Pt layer is excited by both the optical pulse and the hot electrons [33], the transfer of the hot electron angular momentum could be sufficient to determine the final state of the magnetization of [Co/Pt] as it cools. Thus it is expected that the longer time spin-polarized currents that are parallel to the Gd sublattice will determine the final state. At the same time, GdFeCo layer is also reversed as a result of the AO-HIS mechanism. Thus even though the [Co/Pt] layer final state magnetization is determined by the Gd sublattice, the FeCo sublattice and the [Co/Pt] magnetization are always parallel in the final state. This mechanism might be also contribute to the recent single shot AOS demonstrated in Co/Gd bilayer system [35], in which non-local transfer of angular momentum might reverse the magnetization of Co since the exchange interaction between Gd and Co exists only at the interface.

To explore the validity of the above spin polarized hot electron transport model to explain the [Co/Pt] magnetization switching, we have grown several spin-valve structures where we have modified the Cu interlayer. First GdFeCo / Cu / [Co/Pt] spin-valve have been grown with different Cu thicknesses ranging from 5 nm to 80 nm. Single-pulse [Co/Pt] switching could be observed up to 30-nm Cu interlayers. The loss of the [Co/Pt] reversal is attributed to the limited spin-diffusion length of the hot electrons in the Cu layer, estimated to 13 nm by Schellekens *et al.*[26]. On other way to reduce the spin polarization of the hot electrons consists to insert 1 to 5 nm Pt layer in the Cu spacer layer (shown schematically in Fig. 5a). Pt has a significantly shorter spin diffusion length than Cu and is on the order of a few nanometers [26]. While the average magnetic properties of samples with and without Pt layer are very similar (see magnetization curves in supplementary information, sec. E) it is expected that the Pt layers will depolarize the hot electrons before reaching the [Co/Pt] layer. This should limit the ability to deterministically switch the [Co/Pt] layer. Figure 5b shows the MOKE images after the irradiation of fs-laser pulses for the samples with different Pt thicknesses $t_{Pt}$. For $t_{Pt}$ = 0 nm we see the P+ to P- and back to P+ switching as described earlier and the [Co/Pt] layer is 100% switched. With increasing Pt thickness, the magnetic contrast around center of the spot gradually decreases indicating the samples are transitioning from AO-HIS to demagnetizing with increasing Pt. By analyzing the magnetic contrast values in details, we plot in Fig. 5c the change in magnetization of the [Co/Pt] layer with increasing Pt thickness. We see that with increasing Pt thickness the samples transition from deterministic switching to demagnetization where the GdFeCo and [Co/Pt] layers break into small domains. The decay length obtained by exponential fitting was ~ 2 nm as shown in Fig. 5c (See detail of the analysis in supplementary information, sec. F). This decay length is consistent with a previous report for the spin diffusion length of hot electron in Pt layer (3 nm) [23,26] and is quite a bit shorter than penetration depth of light in Pt layer ~ 13 nm calculated from imaginary part of the refractive index of 2.85 + $i$ 4.96 for the wavelength of 800 nm. Thus, the magnetization of [Co/Pt] after irradiation of fs-laser pulse changes with increasing Pt thickness more drastically than the change in light absorption and temperature. Moreover, we also performed light absorption calculation. It was found that total light absorption (i.e., temperature rising) in the sample was not changed significantly with different Pt thickness, (see supplementary information, Sec. G). Based on these studies, we conclude that magnetization switching of the GdFeCo / Cu / [Co/Pt] is mediated by spin-polarized hot electron transport.

*Conclusion*

In this study, we demonstrated that we could access the four remanent magnetic configurations in GdFeCo / Cu / [Co/Pt] spin-valve structures, without applied field using single femto-second pulses. After studying the effect of the GdFeCo concentration and of the spacer layer we concluded that the final state of the magnetization switching of [Co/Pt] is mediated by spin-polarized hot electron transport. The final state is consistent with the expected spin polarization being parallel to Gd moment in the GdFeCo layer due to the slower demagnetization of Gd compared to the one of the FeCo spins. These hot spin-polarized electrons transfer their angular momenta which, in combination with optical heating, are able to deterministically switch the magnetization of [Co/Pt]. This conclusion is supported by inserting Pt layers inside the Cu spacer layer of the spin-valve to depolarize the optically-induced spin current resulting in the thermal demagnetization of the [Co/Pt] layer. This work provides a new approach to deterministically single-pulse switch ferromagnetic layers and a pathway to engineering materials for opto-magnetic multi-bit recording using spin-valve structures.

Methods

1. Sample preparation

All samples were prepared by physical vapor deposition. Base pressure used to deposit multilayer film was about $1\times 10^{-7}$ Torr. Basic stacking structures used in this study for the spin-valve structure are as follows,

Glass sub. / Ta (5) / [Pt(1)/Co(0.6)]$_4$ / Cu ($t_{Cu}$) / Gd$_x$(Fe$_{87}$Co$_{13}$)$_{1-x}$ (5) / Ta (5), (thickness in nm)

Cu thickness was varied from 5 nm to 80 nm. Gd composition was varied from 21.9% to 27.2%.

Stacking structures of the reference sample for the Pt layer insertion are as follows,

Glass sub. / Ta (5) / [Pt(1)/Co(0.6)]$_4$ / Cu (7) / Pt ($t_{Pt}$) / Cu (7) / Gd$_{23.3}$Fe$_{66.7}$Co$_{10}$ (5) / Ta(5).

2. AOS measurement

Ti: sapphire fs laser source and regenerative amplifier are used for the pump laser beam in AOS measurement. Wavelength, pulse duration and repetition rate of the fs laser are 800 nm, 35 fs, and 5 kHz, respectively. The $1/e^2$ spot size $2w_0$ is ~50 μm. No external magnetic field is applied during measurement. Four different magnetic configurations are realized by using permanent magnet before taking images. MOKE images were obtained from the other side of the film. LED light source with the wavelength of 628 nm was used for taking MOKE images.

3. Data analysis

The movies of MOKE images during irradiation of fs-laser pulses were taken by CCD camera. After taking movies, the images of each slice were subtracted from the initial slice of the movies to exclude the background of the images. And then, average contrast values of the initial images were added to each subtracted images. The median and mean filters were applied to reduce noise in the images. The gray scale magnetic contrast values were converted to the color (lower contrast is blue, middle contrast is white, and higher contrast is red).

Reference

[1] Stanciu, C. D. *et al*. All-Optical Magnetic Recording with Circularly Polarized Light. *Phys. Rev. Lett*. 99, 047601 (2007)

[2] Radu, I. *et al*. Transient ferromagnetic-like state mediating ultrafast reversal of antiferromagnetically coupled spins. *Nature* **472**, 207 (2011)

[3] Ostler, T. A. *et al*. Ultrafast heating as a sufficient stimulus for magnetization reversal in a ferrimagnet. *Nat. Commun*. **3**, 666 (2012)

[4] Mangin, S. *et al.* Engineered materials for all-optical helicity-dependent magnetic switching. *Nat. Mater.* **13**, 286 (2014)

[5] Lambert, C. -H. *et al.* All-optical control of ferromagnetic thin films and nanostructures. *Science* **345**, 1337 (2014)

[6] Stupakiewicz, A. Szerenos, K., Afanasiev, D., Kirilyuk, A. & Kimel, A. V. Ultrafast photo-magnetic recording in transparent medium. *Nature* **542**, 71 (2017)

[7] Wilson, R. B. *et al.* Ultrafast magnetic switching of GdFeCo with electronic heat currents. *Phys. Rev. B* **95**, 180409(R) (2017)

[8] Xu, Yong *et al.* Ultrafast Magnetization Manupulation Using Single Femtosecond Light and Hot-Electron Pulses. *Adv. Mater.* **29**, 1703474 (2017)

[9] Yang, Yang *et al.* Ultrafast magnetization reversal by picosecond electrical pulses. *Sci. Adv.* **3**, e1603117 (2017)

[10] El Hadri, M. S. *et al*. Two types of all-optical magnetization switching mechanisms using femtosecond laser pulses. *Phys. Rev. B* **94**, 064412 (2016)

[11] Sander, D. *et al.* The 2017 Magnetism Roadmap. *J. Phys. D: Appl. Phys.* **50**, 363001 (2017)

[12] El Hadri, M. S., Hehn, M., Malinowski, G., & Mangin, S. Materials and devices for all-optical helicity-dependent switching. *J. Phys. D: Appl. Phys.* **50**, 133002 (2017)

[13] Takahashi, Y. K. *et al.* Accumulative Magnetic Switching of Ultrahigh-Density Recording Media by Circularly Polarized Light. *Phys. Rev. Appl.* **6**, 054004 (2016)

[14] Ellis, M. O. A., Fullerton, E. E. & Chantrell, R. W. All-optical switching in granular ferromagnets caused by magnetic circular dichroism. *Sci. Rep.* **6**, 30522 (2016)

[15] El Hadri, M. S. *et al.* Domain size criterion for the observation of all-optical helicity-dependent switching in magnetic thin films. *Phys. Rev. B* **94**, 064419 (2016)

[16] Gorchon, J. *et al.* Model for multishot all-thermal all-optical switching in ferromagnets. *Phys. Rev. B* **94**, 020409(R) (2016)

[17] Bergeard, N. *et al.* Ultrafast angular momentum transfer in multisublattice ferrimagnets. *Nat. Commun.* **5**, 3466 (2014)

[18] Graves, C. E. *et al.* Nanoscale spin reversal by non-local angular momentum transfer following ultrafast laser excitation in ferrimagnetic GdFeCo. *Nat. Mater.* **12**, 293 (2013)

[19] Wienholdt, S., Hinzke, D., Carva, K., Oppeneer, P. M., & Nowak, U. Orbital-resolved spin model for thermal magnetization switching in rare-earth-based ferrimagnets. *Phys. Rev. B* **88**, 020406(R) (2013)

[20] Mentink, J. H. *et al.* Ultrafast Spin Dynamics in Multisublattice Magnets. *Phys. Rev. Lett.* **108**, 057202 (2012)

[21] Lavrijsen, R. *et al.* Magnetic ratchet for three-dimensional spintronic memory and logic. *Nature* **493**, 647 (2013)

[22] Zhang, S. *et al.* Three dimensional magnetic abacus memory. *Sci. Rep.* **4**, 6109 (2014)

[23] Suto, H. *et al.* Three-dimensional magnetic recording using ferromagnetic resonance. *Jpn. J.*

*Appl. Phys.* **55**, 119204 (2016)

[24] Malinowski, G. *et al.* Control of speed and efficiency of ultrafast demagnetization by direct transfer of spin angular momentum. *Nat. Phys.* **4**, 855 (2008)

[25] Rudolf, D. *et al.* Ultrafast magnetization enhancement in metallic multilayers driven by superdiffusive spin current. *Nat. Commun.* **3**, 1037 (2012)

[26] Schellekens, A. J., Kuiper, K. C., de Wit, R. R. J. C., & Koopmans, B. Ultrafast spin-transfer torque driven by femtosecond pulsed-laser excitation. *Nat. Commun.* **5**, 4333 (2014)

[27] Choi, G. -M., Min, B. -C., Lee, K. -J., & Cahill, D. G. Spin current generated by thermally driven ultrafast demagnetization. *Nat. Commun.* **10**, 1038 (2014)

[28] Choi, G. -M., Moon, C. -H., Min, B. -C., Lee, K. -J., & Cahill, D. G. Thermal spin-transfer torque driven by the spin-dependent Seebeck effect in metallic spin-valves. *Nat. Phys.* **11**, 576 (2015)

[29] Choi, Gyung-Min, & Min, Byoung-Chul. Laser-driven spin generation in the conduction bands of ferrimagnetic metals. *Phys. Rev. B* **97**, 014410 (2018)

[30] Vahaplar, K. *et al.* All-optical magnetization reversal by circularly polarized laser pulses: Experiment and multiscale modeling. *Phys. Rev. B* **85**, 104402 (2012)

[31] Gorchon, J. *et al.* Role of electron and phonon temperatures in the helicity-independent all-optical switching of GdFeCo. *Phys. Rev. B* **94**, 184406 (2016)

[32] Kirilyuk, A., Kimel, A. V., & Rasing, T. Laser-induced magnetization dynamics and reversal in ferrimagnetic alloys. *Rep. Prog. Phys.* **76**, 026501 (2013)

Demagnetization induced by hot-electron

[33] Bergeard, N. *et al*. Hot-Electron-Induced Ultrafast Demagnetization in Co/Pt Multilayers. *Phys. Rev. Lett*. **117**, 147203 (2016)

[34] Gorchon, J. *et al.* Single shot ultrafast all optical magnetization switching of ferromagnetic Co/Pt multilayers. *Appl. Phys. Lett.* **111**, 042401 (2017)

[35] Lalieu, M. L. M., Peeters, M. J. G., Haenen, S. R. R., Lavrijsen, R. & Koopmans, B. Deterministic all-optical switching of synthetic ferrimagnets using single femtosecond laser pulses. *Phys. Rev. B* **96**, 220411(R) (2017)

Acknowledgement

S. I. would like to thank the Japan Society for the Promotion of Science (JSPS) for a Grant-in-Aid for JSPS Fellows (28-7881). We would like to thank Crosby Chang and Stephane Suire for their assistance of SQUID-VSM measurement. This work was supported by the ANR-NSF Project, ANR-13-IS04-0008-01, COMAG, ANR- 15-CE24-0009 UMAMI and by the ANR-Labcom Project LSTNM, by the Institut Carnot ICEEL for the project « Optic-switch » and Matelas and by the French PIA project 'Lorraine Université d'Excellence', reference ANR-15-IDEX-04-LUE. Experiments were performed using equipment from the TUBE. Davm funded by FEDER (EU), ANR, Région Grand Est and Metropole Grand Nancy.

Author contributions

G. M., M. H., E. E. F., and S. M. conceived the research topic. S. I. and Y. X. performed sample fabrication, M. H. helped for sample preparation. M. D. and G. M., developed AOS measurement set-up. S. I. performed AOS experiment and data analysis. J. G. performed light absorption calculation. S. I., J. G., G.M., E.E.F and S. M. wrote manuscript. All authors contributed to discuss the measurement results.

Figures

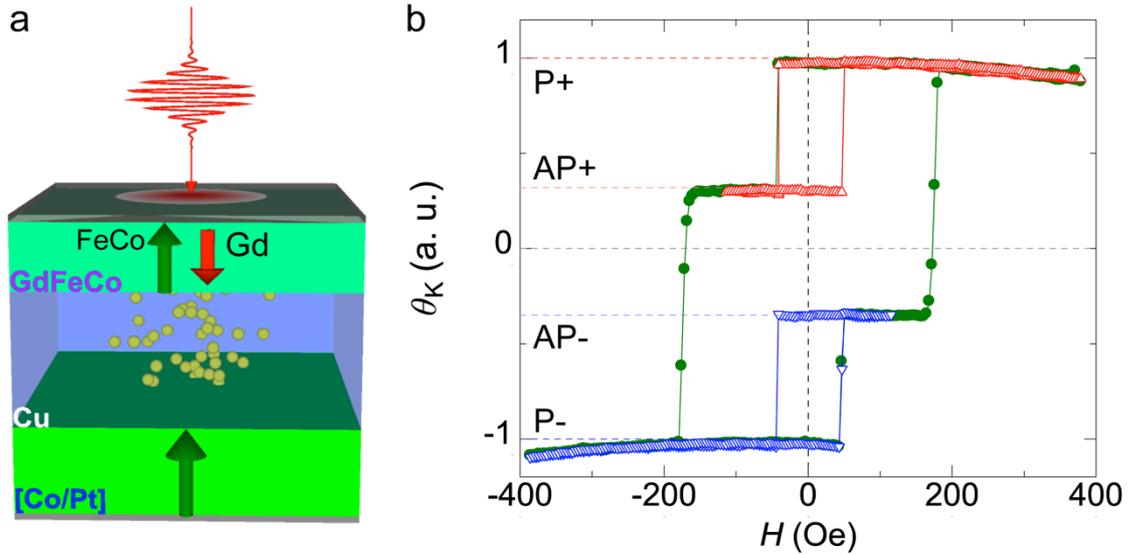

**Figure 1 Properties of the GdFeCo / Cu / [Co/Pt] spin valve structure.** a) Schematic representation of the Glass sub. / Ta (5 nm) / $Gd_{23.3}(FeCo)_{76.7}$ (5 nm) / Cu(9.3 nm) / [Co(0.6 nm)/Pt(1.0 nm)]$_4$ / Ta (5 nm) sample where the GdFeCo and [Co/Pt] magnetic layers have perpendicular anisotropy and are separated by a 9.3 nm thick Cu layer. b) Normalized magneto-optic Kerr rotation ($\theta_\kappa$) as a function of the magnetic field (H) applied perpendicularly to the film plane. Red and blue open symbols are minor loops corresponding to the magnetization reversal of GdFeCo which are perfectly centered around the zero field axis.

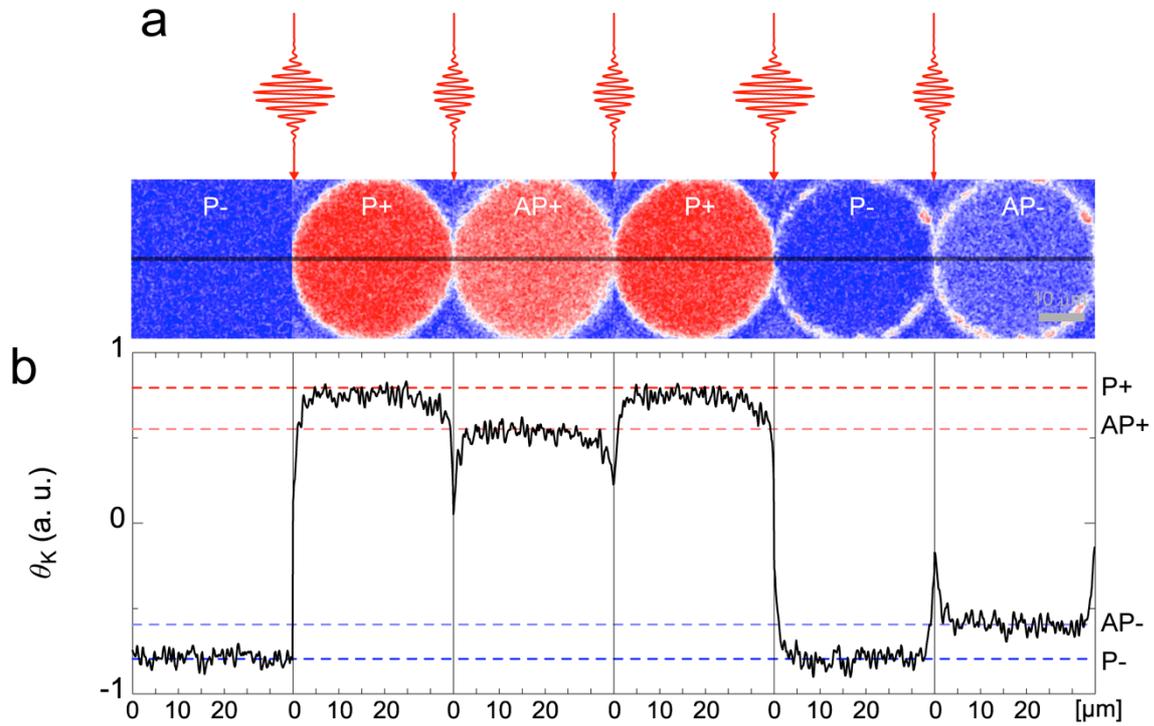

**Figure 2 Multi-level switching of GdFeCo /Cu/[Co/Pt] spin valve structure using 35-fs single light pulses.** a) Experimental demonstration of magnetic configurations obtained consecutively after single optical pulses. Starting from a saturated state P- (resp. P+) a single intense pulse (0.5 μJ) induce a switching into the P+ state (resp. P-). A single moderate light pulse (0.2 μJ) induces a transition from a P- state (resp. P+) to an AP- (resp. AP+) state. All the measurements demonstrate that the GdFeCo layer switching can be obtained using a single moderate light pulse (0.2 μJ) whereas the complete switching of both layer is obtained for single intense pulse (0.5 μJ). b) Normalized averaged magnetic contrast obtained along the black line shown in a), averaged within the width of 5 μm. The four different levels allow to quantitatively define the four magnetic states (P+, P-, AP+, AP-) which can be reached using a sequence of single 35-fs laser pulses.

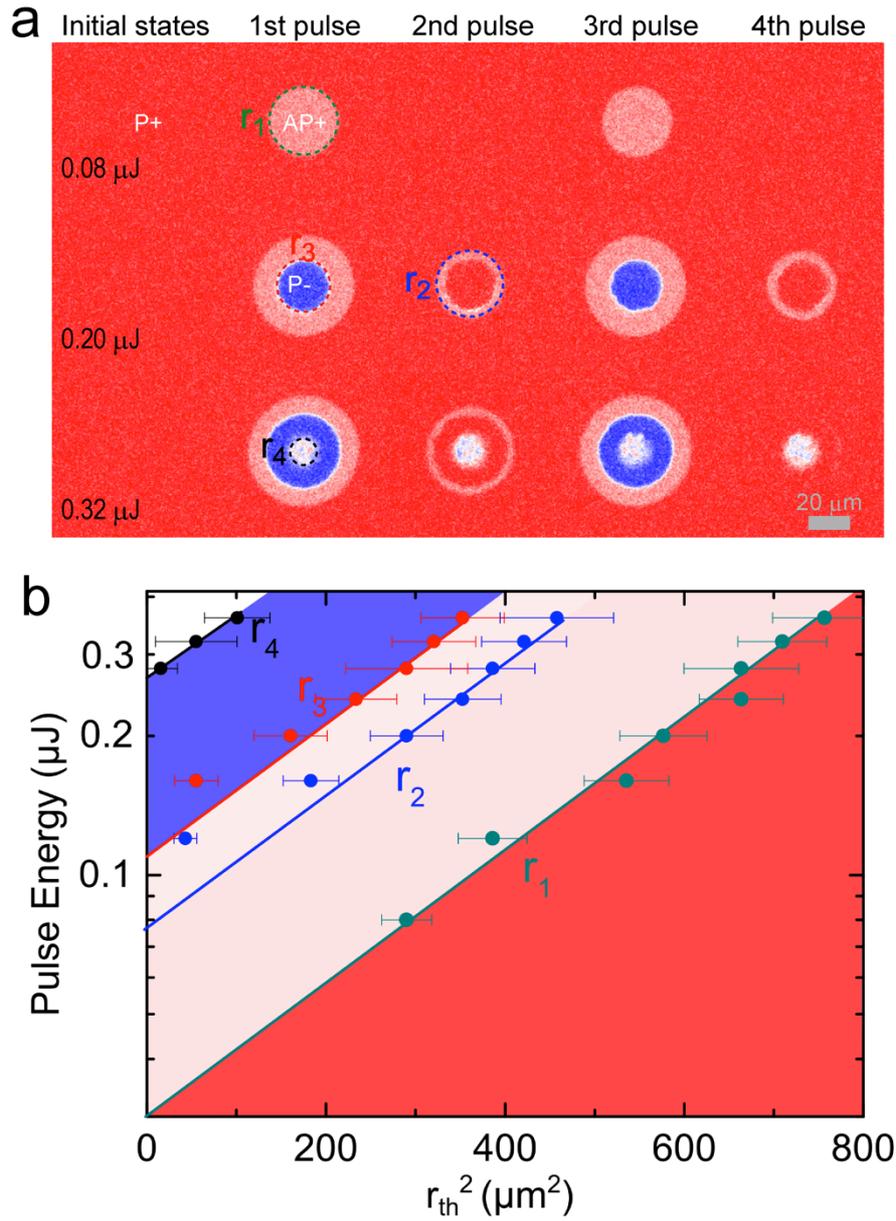

**Figure 2 Single pulse all-optical switching as a function of pump energy for $Gd_{23.3}(FeCo)_{76.7}$ / Cu(9.3) / [Co/Pt]$_4$ sample.** a) MOKE images obtained after four consecutive 35-fs pulses of various pump energy (the time between two pulses is 2 seconds). The same experiments has been repeated with different pump energy ranging from 0.08 to 0.36 μJ. b) Threshold radius ($r_i$) such that for r> $r_1$: there is no light effect, $r_1$>r>$r_2$ AO-HIS single pulse GdFeCo layer reversal. $r_2$>r>$r_3$ Starting from P+ the GdFeCo magnetization is switched once to reach an AP+ configuration and then light pulses have no effect; $r_3$>r>$r_4$: AO-HIS is observed for both layers; $r_4$ >r the energy is too high and the all stack demagnetized.

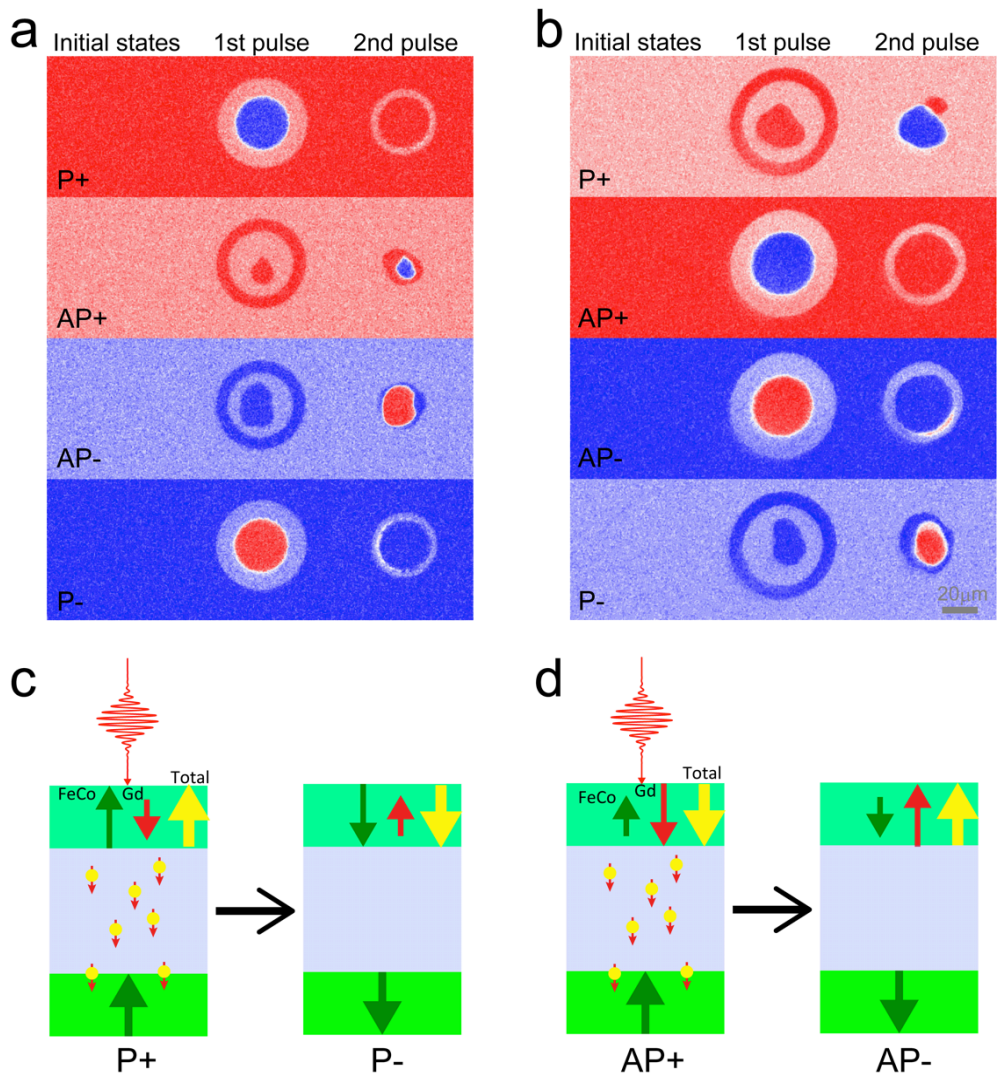

**Figure 3 All-Optical Switching in GdFeCo / Cu / [Co/Pt] spin-valves with different GdFeCo concentration.** a) AOS results for FeCo-rich $Gd_{23.3}(FeCo)_{76.7}$ (5 nm)/ Cu(9.3 nm) / $[Co/Pt]_4$ sample. b) AOS results for Gd-rich $Gd_{26.4}(FeCo)_{73.6}$ (5 nm) / Cu(10 nm) / $[Co/Pt]_4$ sample. Since MOKE is more sensitive to the FeCo magnetic sublattice, the magnetic contrast signals of AP+ states are larger than that of P+ state in the case of spin-valve sample with Gd-rich GdFeCo. Schematic illustration of spin-polarized hot electron transport induced magnetization switching for c), FeCo-rich and d), Gd-rich GdFeCo / Cu / [Co/Pt] spin valve structures.

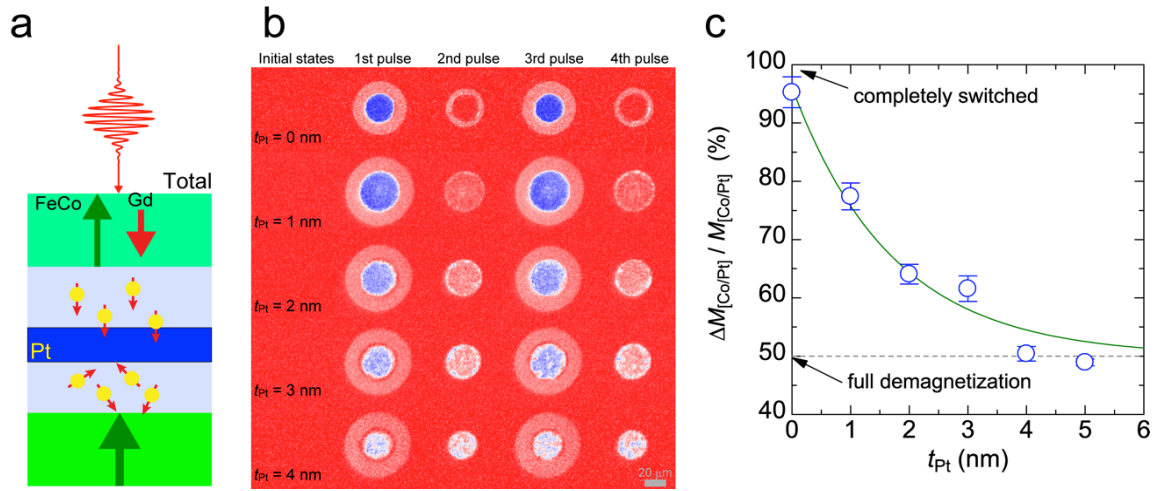

**Figure 4 Evolution of the Co/Pt magnetization switching on Pt insertion layers in Cu.** a) Schematic illustration of the role of the Pt spin scattering layer inserted in the Cu spacer layer. b) MOKE images after irradiation of consecutive single fs-laser pulses on GdFeCo / Cu (7) / Pt ($t_{Pt}$) / Cu (7) / [Co/Pt] samples and GdFeCo / Cu (15) / [Co/Pt] sample ($t_{Pt}$ = 0 nm). c) The changes in magnetization of [Co/Pt] around the center of the spot estimated from magnetic contrast are plotted as a function of Pt thicknesses. 100 % indicates completely switched and 50 % indicates full demagnetization. Solid curve is an exponential fitting and decay length obtained from fitting is ~ 2 nm.